\documentclass[dvips]{article}
\usepackage{icrctc07}

\title{\boldmath TenTen: A New Array of Multi-TeV Imaging Cherenkov Telescopes}
\shorttitle{TenTen: A New Multi-TeV Array}
\authors{G.~Rowell, V.~Stamatescu, R.~Clay, B.~Dawson, J.~Denman, R.~Protheroe, A.~Smith, G.~Thornton, N.~Wild}
\shortauthors{G. Rowell et al.}
\afiliations{School of Chemistry \& Physics, University of Adelaide, Adelaide 5005, Australia}

\abstract{
  The exciting results from H.E.S.S. point to a new population of $\gamma$-ray sources at
  energies E$>$10~TeV, paving the way for future studies and new discoveries in the multi-TeV energy range. Connected
  with these energies is the search for sources of PeV cosmic-rays (CRs) and the study of multi-TeV $\gamma$-ray
  production in a growing number of astrophysical environments. {\em TenTen} is a proposed stereoscopic
  array (with a suggested site in Australia) of modest-sized (10 to 30m$^2$) Cherenkov imaging telescopes with a wide field of view 
  (8$^\circ$ to 10$^\circ$ diameter) optimised for the E$\sim$10 to 100~TeV range. {\em TenTen} will achieve an effective 
  area of $\sim$10~km$^2$ at energies above 10~TeV. We outline here the motivation for {\em TenTen} and summarise key
  performance parameters.
}

\email{growell@physics.adelaide.edu.au}

\begin{document}
\maketitle
%Begin the section.

\section{\boldmath Motivation for Multi-TeV Studies}

Ground-based $\gamma$-ray astronomy operating in the $\sim$0.1 to $\sim$10~TeV range has become a mainstream
astronomical discipline due to the exciting results from H.E.S.S. \cite{HESS} in the
Southern Hemisphere over recent years. In the Northern Hemisphere the MAGIC \cite{MAGIC} and MILAGRO \cite{MILAGRO} telescopes
are now producing results and similar high impact can soon be expected from VERITAS \cite{VERITAS}.
The TeV source catalogue extends to over 30 individual sources, and we are now able to perform
detailed studies of extreme environments capable of CR particle acceleration. 
Several key points can be extracted from the results of H.E.S.S. and others which motivate development
of a new dedicated instrument for studies at multi-TeV ($E>$few~TeV) energies (see also \cite{Rowell:2,Felix:1}):\\

\begin{figure}[t]
  \centering
  \includegraphics[width=0.45\textwidth]{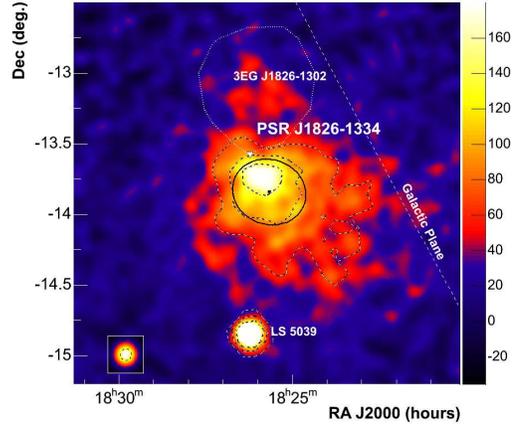}
  \caption{H.E.S.S. image of the PWN HESS~J1825$-$137 \cite{HESS-HESSJ1825} resulting from 52~hr observations. A considerably
    weaker source is revealed to the north of HESS~J1825$-$137, 
    and is possibly a counterpart to the EGRET source 3EG~J1826$-$1302. This weaker source is unlikely to be 
    studied further in depth by H.E.S.S. This image nicely illustrates a variety of Galactic source types in the same FoV --- 
    a strong extended PWN; weak extended source with MeV/GeV counterpart and a pointlike compact binary (LS~5039).}
  \label{fig:j1825}       
\end{figure}

{\bf Increasing variety of TeV Sources:} The number of environments established as sources of gamma radiation (to 
energies exceeding $\sim 10$~TeV in many cases) is growing and include, in the case of Galactic sources --- shell-type 
supernova remnants (SNRs), pulsar-wind-nebulae (PWN), compact binaries and/or X-ray binaries, young stellar systems/clusters
and molecular clouds acting as targets for CRs in their vicinity. There are also several Galactic sources (all extended) 
that to-date have no known counterpart at lower energies, and so remain unidentified.
Extragalactic sources comprise active galaxies with jets aligned along our line of sight, the so-called Blazars,
and at least one misaligned Blazar.\\
{\bf Hard Photon Spectra:} The majority of new Galactic sources generally exhibit hard power law photon spectra $dN/dE \sim E^{-\Gamma}$ 
where $\Gamma<2.5$ without indication of cutoffs, suggesting that their emission extends beyond 10~TeV.\\
{\bf Extended Sources \& Large Field of View:} The majority of Galactic sources are extended (up to several degrees in scale) 
in morphology, providing insight into $\gamma$-ray production and transport processes, especially when coupled with multiwavelength 
results. Effective studies of these sources therefore require large instantaneous fields of view, not only to encompass sources of 
interest but to also allow adequate selection of regions for CR background estimation.  The 5$^\circ$ FoV cameras in H.E.S.S. 
for example have been designed with these aspects in mind. This large FoV has permitted highly successful surveys of the inner Southern 
Galactic Plane within just a few years \cite{HESS-ScanI,HESS-ScanII}, and also the establishment of degree-scale morphology in several
strong sources.\\
{\bf Limited Multi-TeV Sensitivity of Present Instruments:} Current intruments operate with a $\sim$0.1~TeV threshold energy and 
effective collection area ($E>10$~TeV) of less than 1~km$^2$. The majority of source studies are therefore made in the 0.1 to 
$\sim$10~TeV band. The fluxes of these new sources are in the few to $\sim$15\%~Crab flux range, reflecting the intrumental sensitivities.
With the limited observational opportunities due to the ever growing source catalogues, 
the accumulation of high statistics in the multi-TeV band is difficult. Fig.~\ref{fig:j1825} illustrates this point
where a new weak TeV source is revealed to the north of HESS~J1825$-$137 after only deep ($>50$~hr) observations. 
Further, detailed studies of this weak source, motivated perhaps by its relatively rare coincidence with a MeV/GeV EGRET source, 
would not be practical with H.E.S.S. Such studies would require $>$100~hr of observations and face stiff competition from other 
source programmes. \\  
{\bf \boldmath $E>10~TeV$ Sources Already Exist:} For those Galactic sources (two shell-type SNRs and several PWN 
\cite{HESS-MSH,HESS-VelaX,HESS_VelaJnrII,HESS-HESSJ1825,HESS_RXJ1713III}) with 
strong fluxes ($>15$\%~Crab) and/or deep observation times ($\geq 50$~hr), photon spectra have been established to energies 
$\sim$50~TeV or greater, demonstrating that particle acceleration to energies exceeding 100~TeV is occurring in these types 
of objects. MILAGRO has also recently revealed degree-scale emission (with total flux exceeding 1~Crab) at energies above 10~TeV in the 
Cygnus and other regions of the Northern Galactic Plane \cite{Abdo:1,Abdo:2}, highlighting the potential of future
all-sky-monitors in the multi-TeV range.
These points provide clear observational evidence that rich astrophysics potential awaits in the $E>10$~TeV range.

In addition there are strong theoretical grounds for pushing deep into the multi-TeV domain:\\
{\bf Particle Acceleration to the {\em knee} and beyond:}
The desire to understand particle acceleration to the CR {\em knee} ($E\sim 1$~PeV) energy and beyond remains a key motivation for
multi-TeV studies. While it is generally accepted that CRs can be
accelerated in shell-type SNRs \cite{Ginzburg:1} to energies $E_{\rm max}\sim$few$\times 10^{14}$~eV \cite{Lagage:1} 
(via the diffusive shock acceleration process), there is considerable uncertainty as to 
how particles can reach the {\em knee} energy and beyond (eg. \cite{Kirk:1,Drury:1}) in so-called {\em Pevatrons}. 
Several ideas have been put forward, for example: strong amplification of pre-shock magnetic fields \cite{Bell:1}; local 
Gamma-Ray-Bursts (GRBs) \cite{Waxman:1,Dermer:1}; and superbubbles which combine the effects of many SNRs and maybe 
Wolf-Rayet/OB stellar winds \cite{Drury:2,Bykov:1,Parizot:1,Binns:1}. 
Extragalactic sources with large-scale kpc shocks such as galaxy clusters (eg. \cite{Volk:1}) and AGN jets and giant lobes 
(eg. \cite{Hillas:1}) could also be sources.
Only observations around 100~TeV and greater can begin to solve the mystery of PeV CR acceleration.\\ 
{\bf \boldmath $E>10$~TeV --- Easier Separation of Hadronic \& Electronic Components:} 
A major complication in interpreting present results in the 0.1 to $\sim$10~TeV range concerns the separation of $\gamma$-ray
components from accelerated hadrons (from secondary $\pi^\circ$-decay) and those from accelerated electrons (most commonly from 
inverse-Compton scattering). Multiwavelength information, in particular
at radio and X-ray energies, can provide constraints on these components but often one requires model-dependent assumptions
to decide the nature of the parent particles. At energies $E>$10~TeV the electronic component can be suppressed due to strong
radiative synchrotron energy losses suffered by electrons in magnetised post-shock environments, such as that in shell-type SNRs. 
In addition, the Klein-Nishina effect on the inverse-Compton cross-section can significantly reduce the efficiency of this process.
Except in those cases where a strong source of electrons exists, such as in PWN, interpretation of $E>10$~TeV spectra may therefore be 
much more confidently interpreted as arising from accelerated hadrons.\\
{\bf \boldmath Probing Local Intergalactic/Interstellar Photon Fields:} $E>10$~TeV photons can indirectly
probe ambient soft photon fields. In the $\sim$10 to $\sim$100~TeV energy range, 
absorption on the cosmic infra-red background (CIB) in the 10 to 100~$\mu$m range dominates with mean free paths extending beyond 1~Mpc. 
Constraints on the (nearby) intergalactic CIB via $\gamma$-ray spectral studies of nearby extragalactic sources, such as M~87 (an 
established TeV source) can yield important information concerning star and galaxy formation in our local intergalactic 
neighbourhood \cite{Felix_book}. Constraints on the
interstellar CIB may also be possible via $E>10$~TeV spectral studies of Galactic source populations \cite{Moskalenko:1}. 

\section{TenTen: Initial Simulation Study \& Performance}

Given that source fluxes rapidly decrease with energy (typically via a power law), any dedicated instrument operating in 
the multi-TeV energy domain would need to achieve a very large effective collection area $A_{\rm eff} \sim$10~km$^2$. 
While there are several promising ways to achieve 10~km$^2$ using ground-based techniques, earlier simulations 
\cite{Plya:1} have shown that a proven, technically straightforward, and yet highly sensitive method would employ  
stereoscopy in an array of 30 to 50 modest-sized imaging atmospheric Cherenkov telescopes (IACTs) in a cell-based approach.
Each telescope would have mirror area 10 to 30~m$^2$, field of view (FoV) 5$^\circ$ to 10$^\circ$, and inter-telescope 
spacing within a single cell $\geq$200 metres (in contrast to $\sim$100~m employed by arrays such as H.E.S.S.). 
The large FoV, limited practically by optical aberrations, allows events to trigger out to core distances $\geq$200~m, thereby 
increasing the effective collection area of a cell. We propose here such an array, known as {\em TenTen}, which stands for 
10~km$^2$ above 10~TeV. Similar and other ideas for 100~TeV studies have also been suggested 
\cite{Kifune-AllSkyII,Yoshikoshi:1,LeBohec:1}.

\begin{figure*}[t]
  \centering
  \hbox{
    \begin{minipage}{0.65\textwidth}
      \centering
      \includegraphics[width=\textwidth]{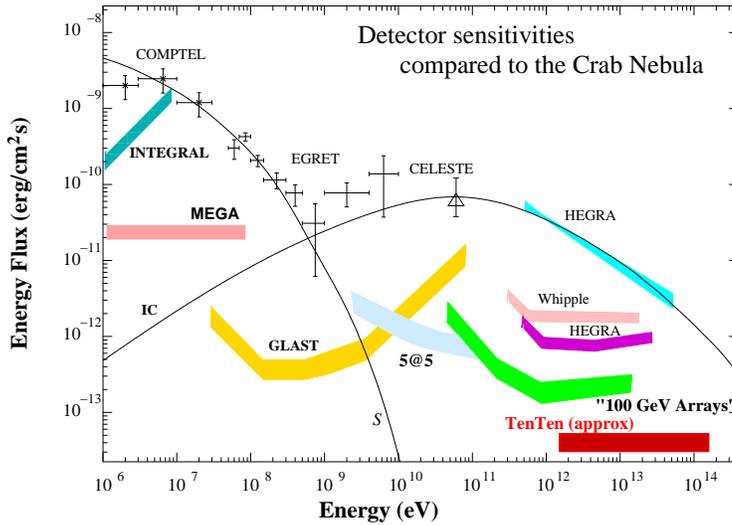}
    \end{minipage}
    \begin{minipage}{0.32\textwidth}
      \caption{Approximate point source energy flux sensitivity for {\em TenTen} 
	(assuming 10 cells of 5 telescopes discussed in text)
	in comparison with other instruments in the $\gamma$-ray regime and the flux from the Crab Nebula. 
	An observation time of 50~hr and signal significance of 5$\sigma$ is demanded. The {\em TenTen} sensitivity is 
	estimated from its collection area improvement over H.E.S.S.  (similar to ``100 GeV arrays'') 
	from our initial simulation study.} 
      \label{fig:sens}       
    \end{minipage}
  }
\end{figure*}

Our initial simulation study \cite{Rowell_Adelaide} examined the performance of a 
single cell of 5 telescopes, each with mirror area 23.8~m$^2$ (84x60~cm spherical mirror facets) and 1024 pixel camera 
spanning 8.2$^\circ$ diameter (with pixel diameter 0.25$^\circ$). The layout of the cell 
has the outer four telescopes arranged in a square of side length $L$ with a single telescope at the centre 
(similar to the HEGRA IACT-System layout\cite{HEGRA}). Gamma-ray and proton extensive air shower simulations (30$^\circ$ zenith ---
with CORSIKA v6.204 \cite{CORSIKA} and SIBYLL \cite{SIBYLL}) coupled with telescope responses 
(based on \cite{simtelarray}) were used to investigate  
basic performance parameters of the cell. An observation altitude 200~m a.s.l. was chosen since we are investigating sites
in Australia. For $E>10$~TeV, low-altitude sites may be beneficial compared to mid/high altitudes
due to the larger distances between telescopes and shower maxima. Details concerning this study including ongoing work 
are summarised in our companion paper/poster \cite{Stam:1}. Briefly, we found that for a cell with side length $L=300$~m,  
an on-axis $\gamma$-ray effective collection area $A_{\rm eff}$ exceeding 1~km$^2$ for $E>30$~TeV can be achieved in a 
single cell, 
and that somewhat similar cosmic-ray background rejection power and arc-minute angular resolution is achieved in 
the $E>10$~TeV range as H.E.S.S. and the HEGRA IACT-System achieve(d) in their respective energy ranges. An energy threshold in the 1
to few~TeV range is also indicated.
This encouraging result suggests that expanding the array to (for example) $\sim$10 sufficiently spaced cells (so that there are no
common events between cells) could yield collection 
areas $\sim$10~km$^2$, exceeding that of H.E.S.S. by factors approaching 50 at 100~TeV.  The approximate 
flux sensitivity (based on the improvement in collection area over H.E.S.S.) and energy coverage of {\em TenTen} is depicted in 
Fig.~\ref{fig:sens}.
Flux sensitivities for large extended sources such as the 2$^\circ$ diameter shell type SNR RX~J0852.0$-$4622 would be 
$\sim 10^{-12}$~erg cm$^{-2}$ s$^{-1}$ at 10~TeV. For point-like sources, the accessible 
fluxes would be a factor of 10 to 20 lower again (less than $\sim 10^{-13}$~erg cm$^{-2}$ s$^{-1}$ above 10~TeV). 
Importantly, a 1 to a few~TeV energy threshold would also allow detailed studies of 
existing $\gamma$-ray sources that are at the threshold of detection for H.E.S.S. Despite our simulations being limited to
$E\leq100$~TeV, we also expect high collection area above this energy. Note that in the 200~TeV to 100~PeV range, $\gamma$-ray 
absorption on the cosmic microwave background (CMB) becomes important, with mean free path $\leq$100~kpc \cite{Coppi:1}, limiting
the focus of $\gamma$-ray astronomy in this energy range to Galactic sources.

\section{Conclusions}

We have outlined the motivation for a new array of IACTs achieving 10~km$^2$ at $E>10$~TeV and described some 
important performance parameters. This array, known as {\em TenTen}, could also be considered complementary to 
future MeV to TeV $\gamma$-ray instruments such as GLAST, HESS-II and MAGIC-II.   
% and HAWC \cite{HAWC}.
Studies are currently underway to further optimise individual telescopes (optics, electronics, camera design), overall layout parameters,
and site potential in Australia.

%{\renewcommand{\baselinestretch}{-0.5}
\scriptsize
%Thanks a lot!}

%\nocite{ref4}
%\nocite{ref5}
%\nocite{ref6}
%\nocite{ref7}
%This is the reference to .bib file (Whitout .bib!)

\bibliographystyle{plain}
\bibliography{icrc0128}

\end{document}